\documentclass[pra,twocolumn,floatfix,notitlepage,groupedaddress]{revtex4-1}

\usepackage[hidelinks,colorlinks=false,citecolor=black,linkcolor=black]{hyperref}
\usepackage{mathtools}
\usepackage{subcaption}
\usepackage{amssymb,amsmath,amstext,amsthm}
\usepackage{graphicx}
\usepackage{epstopdf}
\usepackage{color}
\usepackage{bm}
\usepackage{appendix}
\usepackage[T1]{fontenc}
\usepackage{bbm}
\usepackage{latexsym}

\usepackage{algorithm,algorithmic}

\usepackage{float}
\usepackage{verbatim}
\usepackage{lipsum}
\usepackage{braket}
\usepackage{ulem}

\newcommand{\bigzero}{\mbox{\normalfont\Large\bfseries 0}}

\begin{document}

\title{Nonperturbative Zou-Wang-Mandel effect}
\author{T.J.\,Volkoff}
\affiliation{Theoretical Division, Los Alamos National Laboratory, Los Alamos, NM, USA.}
\author{Diego A.R. Dalvit}
\affiliation{Theoretical Division, Los Alamos National Laboratory, Los Alamos, NM, USA.}

\begin{abstract}
The Zou-Wang-Mandel (ZWM) effect is a remarkable consequence of photon indistinguishability and continuous-variable entanglement in which an optical phase shift is imprinted on photonic modes associated with optical paths that that do not pass through the phase shift source.
By bringing the canonical formalism of continuous-variable Gaussian states to bear on the mode-structure of the ZWM experiment, we show that the physical consequence of implementing optical path identity is a renormalization of quadrature squeezing which governs the entanglement of four effective optical modes. 
Nonperturbative expressions for the ZWM interference patterns and normalized first-order coherence function are derived. Generalizations to $\mathcal{H}$-graph states with more than four modes directly follow from the general method used to analyze the minimal example. We show that a ZWM interferometer with a laser-seeded signal mode, which estimates an idler phase shift by detecting photons that did not propagate through the phase shift, exhibits an optimal sensitivity comparable to that of a laser-seeded SU(1,1) interferometer if path identity is implemented with high fidelity.
\end{abstract}
\maketitle

\section{\label{sec:intro}Introduction}
The Zou-Wang-Mandel (ZWM) experiment demonstrates that the photocurrent from the signal modes arising from spontaneous
parametric downconversion (SPDC) from two coherently pumped crystals \cite{mandelbook} can depend on parameters of a quantum channel applied to the first of the partially aligned idler modes \cite{PhysRevLett.67.318,PhysRevA.41.566,PhysRevA.41.1597}. This remarkable consequence of photon indistinguishability and continuous-variable (CV) entanglement, also known in the literature as quantum-induced coherence by path identity, has influenced a wide range of experimental \cite{PhysRevA.90.045803,PhysRevLett.114.053601} and theoretical work in non-linear quantum optical phenomena \cite{RevModPhys.94.025007}. In recent years, this effect has experienced a revival of interest due to advances in quantum imaging \cite{Lemos2014}, sensing \cite{Paterova_2018,Kalashnikov2016}, and high-dimensional entanglement generation \cite{zeil} experiments with undetected photons, which are all based on the original concept proposed in the ZWM experiment. While the ZWM experiment was carried out and analyzed in the setting of low-gain SPDC \cite{PhysRevA.44.4614}, analyses of analogous ZWM experiments in the high-gain regime of SPDC are vital for understanding how the maximum visibility of the interference beween the signal modes deviates from linearity \cite{WISEMAN2000245,Chekhova,BELINSKY1992303,Kolobov_2017}, and how to incorporate quantum-induced coherence by path identity into advances in SU(1,1) interferometry \cite{10.1063/5.0004873,Hudelist2014,Miller2021versatilesuper,Chekhova} and target detection \cite{PhysRevLett.131.033603}. Such advances could improve path identity-based techniques for imaging with undetected photons \cite{Lemos2014,PhysRevA.92.013832} toward quantum sensing with undetected photons below the standard quantum limit \cite{lemos22,Miller2021versatilesuper}. Although the multimode squeezed states analyzed in this work may indeed be interpreted as high-gain SPDC networks defined by patterns of quantum-induced coherence by path identity, we note that CV entangled states generated by coupling single-mode squeezed states from optical parametric oscillators, central to proposals for CV quantum computing \cite{PhysRevA.79.062318,Madsen2022}, can also be combined with path identity operations and analyzed with the methods described here.

\begin{figure}[t]
    \centering
    \includegraphics[scale=1]{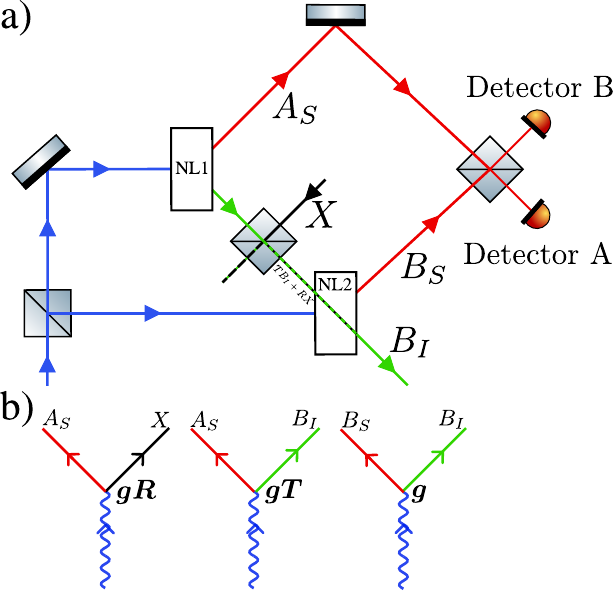}
    \caption{a) The ZWM experiment. The beamsplitter that rotates the $B_{I}$ and $X$ mode is shown only for modeling the imperfect path-identity and is not part of an actual experiment. b) Three vertices contributing to observables of the system of two non-degenerate parametric downconverters in which the idler modes are partially indistinguishable.\footnote{See Ref.\cite{bogo} for perturbative analysis of non-linear quantum optical systems.} Wavy line is the pump mode. The middle process is the source of the renormalized effective squeezing strength.}
    \label{fig:fd}
\end{figure}

In the present work, non-perturbative expressions for the photocurrents and maximum visibility of the signal modes of the ZWM experiment are derived by bringing the canonical formalism of CV Gaussian states to bear on the mode structure of the experiment. The expressions differ from previous analyses of the ZWM experiment in the high-gain SPDC regime \cite{WISEMAN2000245} because we take into account the fact that partial alignment of squeezed CV modes results in a renormalization of the effective squeezing strengths. Specifically, the analyses in Refs.\cite{WISEMAN2000245,Kolobov_2017} treat the ZWM experiment as a sequential pair-production in two downconverters with beamsplitter-coupled idlers, whereas the present analysis treats the two downconverters as a coherent four-mode process with partially distinguishable idler modes. The result is a new set of normal modes for the system, which occur with renormalized squeezing strengths.  We make a quantitative comparison between the result of the present analysis and the result of Ref.\cite{WISEMAN2000245} in terms of the first-order coherence function, which is directly related to the maximal visibility of the interference pattern \cite{mandelbook,agarwal}. Here we note that a completely general non-perturbative description of two non-degenerate parametric downconverters with partially indistinguishable idler modes would take into account non-classical properties of the pump mode. Such a description could be achieved in some parameter regimes by analyzing the quantum multiple three-wave interaction system with non-orthogonal modes using algebraic Bethe ansatz methods \cite{jurco1,JURCO199047,ANDREEV199676,Andreev2001,aa2002}.

\section{Background: squeezed state networks}
The ZWM experiment in its original form is shown in Fig. \ref{fig:fd}a with each labeled mode corresponding to an annihilation operator for a fixed momentum, frequency, and polarization component of the positive frequency part of the quantum electromagnetic field. To make contact with the microscopic description in \cite{PhysRevA.41.1597}, we consider the lossless crystals NL1 and NL2 to be pumped by a monochromatic, constant amplitude laser and that the frequency and momentum matching conditions are satisfied to remove time- and space-dependence. To model possible imperfect alignment of the idler modes from NL1 and NL2, we have explicitly shown a beamsplitter which couples an auxiliary mode $X$ into the $B_{I}$ mode, ideally with $\vert T\vert \approx 1$. Fig.\ref{fig:fd}b shows the Feynman vertices that would occur in a perturbation-theoretic treatment of the present framework, with the middle diagram indicating that pumping NL1 induces coupling between the $A_{S}$ signal mode and the $B_{I}$ idler mode, which is the source of the new normal modes in our analysis. Other sources of time dependence that can arise due to optical path length differences or dephasing between the split pump beams are neglected.  
We use the term SPDC broadly to refer to any process of the form $ab^{\dagger}c^{\dagger} +h.c.$ where different letters indicate orthogonal modes, including e.g., internal degrees of freedom of the electromagnetic field such as polarization modes, as long as the symmetries of the system are respected.
 
We define a squeezed state network, which contains as a special case the $\mathcal{H}$-graph states, i.e., networks of SPDC and single-mode squeezing processes specified by the adjacency matrix of a graph \cite{PhysRevA.76.010302,PhysRevLett.95.133901}. An overview of the canonical description of CV Gaussian systems appears in Appendix \ref{sec:gs}. The squeezed state network $\ket{\Phi_{L}}$ is defined by a Gaussian unitary $U=e^{{1\over 2}a^{\dagger}La^{\dagger \intercal} - h.c.}$ via
\begin{align}
\ket{\Phi_{L}}&:= U\ket{\text{VAC}}=e^{\mathcal{H}}\ket{\text{VAC}}\nonumber \\
\mathcal{H} &= {1\over 2}a^{\dagger}La^{\dagger \intercal} - h.c.
\label{eqn:genstate}
\end{align}
where the $M\times M$ matrix $L$ appearing in the generator $\mathcal{H}$ is complex symmetric with zeroes on the diagonal, and $a^{\dagger}$ is the row vector of creation operators. The $M=2$ case given by $L=g\begin{pmatrix}
0&1\\1&0
\end{pmatrix}$ with $g>0$ is simply a two-mode squeezed state with energy $2\sinh^{2}g$ and perfect positive correlation between the position quadratures of each mode. The larger $M$ cases of $L$ are squeezed state networks, certain cases of which (viz., $\mathcal{H}$-graph states with full-rank adjacency matrix \cite{PhysRevA.83.042335,PhysRevA.76.010302}) correspond to CV cluster states after application of local Gaussian unitaries. In fact, the present paper only considers $\vert L_{i,j}\vert \le g$ for some real, positive squeezing parameter $g$. This restriction describes, e.g., a collection of coherently pumped, identical SPDC elements (the ZWM experiment corresponds to $M=4$). 

The most direct route to the covariance matrix $\Sigma_{\ket{\Phi_{L}}}$ is the Autonne-Takagi diagonalization  of $L$ \cite{horn,PhysRevA.94.062109,PhysRevA.93.062115} (see Appendix \ref{sec:nat} for justification).  Specifically, $L=WDW^{T}$ with $D$ the diagonal matrix consisting of singular values $\lbrace \lambda_{j}\rbrace_{j=1}^{M}$ of $L$, and $W$ an $M\times M$ unitary. The unitary matrix $W$ is associated with a photon number-conserving Gaussian unitary $\mathcal{W}$ defined by its action on the row vector of creation operators as $\mathcal{W}a^{\dagger}\mathcal{W}^{\dagger}=a^{\dagger}W^{\dagger}$, so that 
\begin{align}
\mathcal{W}\ket{\Phi_{L}}&=\mathcal{W}e^{{1\over 2}a^{\dagger}La^{\dagger \intercal} - h.c.}\ket{\text{VAC}}\nonumber \\
&= e^{{1\over 2}a^{\dagger}W^{\dagger}L\bar{W}a^{\dagger \intercal} - h.c.}\ket{\text{VAC}} \nonumber \\
&= {1\over \prod_{j=1}^{M}\sqrt{\cosh \lambda_{j}}}\bigotimes_{j=1}^{M}e^{\tanh \lambda_{j} a_{j}^{\dagger 2}}\ket{0}_{j}
\end{align}
where we used the fact that the vacuum is invariant under number-conserving unitaries and the last line shows a tensor product of single-mode squeezed vacua (see Ref.\cite{PhysRevA.96.023621} for a similar expression for states of massive bosons). It is straightforward to verify that $\Sigma_{\mathcal{W}\ket{\Phi_{L}}}={1\over 2}(e^{-2D}\oplus e^{2D})$. Because $D\succeq 0$, this expression states that a subset of the $\lbrace q_{j}\rbrace_{j=1}^{M}$ quadratures are squeezed and the corresponding subset of the $\lbrace p_{j}\rbrace_{j=1}^{M}$ quadratures are anti-squeezed. Using the isomorphism between the unitary group and the orthogonal symplectic group $U(M)\cong O(2M)\cap Sp(2M,\mathbb{R})$ \cite{Arvind1995}, one obtains that $\mathcal{W}\mathcal{R}\mathcal{W}^{\dagger}=\mathcal{R}\Xi$, with $\Xi:= \begin{pmatrix}\text{Re}W & \text{Im}W \\ -\text{Im}W& \text{Re}W\end{pmatrix}$, so
\begin{align}
\Sigma_{\ket{\Phi_{L}}}&= \Xi \Sigma_{\mathcal{W}\ket{\Phi_{L}}} \Xi^{\intercal}
\label{eqn:phicov}
\end{align}
(see Appendix \ref{sec:gs}). Note that the Autonne-Takagi diagonalization of $L$ can be used to compute the Bloch-Messiah decomposition of the symplectic matrix associated with the Gaussian unitary $U$ \cite{PhysRevA.94.062109}. The Autonne-Takagi diagonalization (and Bloch-Messiah decomposition) are unique up to permutation of the modes \cite{Arvind1995}.

 The exact expression for the expected photon number is obtained from the covariance matrix of a pure, zero-mean Gaussian state via
 \begin{equation}
 \langle a_{j}^{\dagger}a_{j}\rangle_{\ket{\Phi_{L}}}={1\over 2}(\Sigma_{\ket{\Phi_{L}}})_{j,j}+{1\over 2}(\Sigma_{\ket{\Phi_{L}}})_{M+j,M+j}-{1\over 2}.
 \label{eqn:photocurr}
 \end{equation}
 This expression assumes that the idler mode $A_{1}$ is actually created from vacuum by the first parametric downconverter. If the idler mode $A_{1}$ is stimulated by pumping with a laser, then $\ket{\Phi_{L}}$ has a non-zero mean vector, the exact form of which depends on the temporal relation between the pumping of the idler mode and the pumping of the parametric downconverter.
 
 \section{Zou-Wang-Mandel experiment in the high-gain regime}
 
This general formalism can now be applied to the four-mode squeezed state network that occurs in the ZWM experiment. The ZWM state is given by $\ket{\Phi_{L}}$ in (\ref{eqn:genstate}) with 
\begin{equation}
L=g\begin{pmatrix}
  \bigzero
  &  \begin{matrix}
  T & R \\
  1 & 0
  \end{matrix} \\
  \begin{matrix}
  T & 1 \\
  R & 0
  \end{matrix} &
  \bigzero
\end{pmatrix}.
\end{equation}
where we take $T=e^{i\theta_{T}}\vert T\vert$, $0<R<1$, and $\vert T\vert^{2}+R^{2}=1$. With this parameter domain,  $L$ has full rank, but if $R=0$ (i.e., the case of perfect path identity between idler modes) then $L$ has rank 2. Note that $L$ involves only SPDC processes between the modes. Written out explicitly, the generator is
$\mathcal{H}=ga_{A_{S}}^{\dagger}a_{A_{I}}^{\dagger} + ga_{B_{S}}^{\dagger}a_{B_{I}}^{\dagger} -h.c. $, with the mode $A_{I}$ defined by $a_{A_{I}}^{\dagger}:=Ta_{B_{I}}^{\dagger} +Ra_{X}^{\dagger}$.
 We re-enumerate the four physical modes in the experiment by $a_{1}=a_{A_{S}}$, $a_{2}=a_{B_{S}}$, $a_{3}=a_{B_{I}}$, $a_{4}=a_{X}$. Note that the idler mode from the NL1 SPDC is in the two-dimensional subspace spanned by modes $B_{I}$ and an ancillary mode $X$, indicating a general indistinguishability of this mode from $B_{I}$. The mode $X$ also appears in the perturbative theory of the ZWM effect; it must be taken into account because SPDC produces two photons in normalized, orthogonal modes. Perfect alignment of the idlers from the parametric downconverters corresponds to taking $T=1$.  Optical phases associated with the propagation of the signal beams, which are eventually combined to form the modes that are measured at the photocounters, can be introduced at the end of the calculation by acting on the covariance matrix with appropriate orthogonal symplectic matrices in $O(8)\cap Sp(8,\mathbb{R})$. Note that the squeezed state network $\ket{\Phi_{L}}$ is not defined by a product of SPDC operations at NL1 and NL2 occurring, respectively, before and after a beamsplitter. The dynamics defined by the generator $\mathcal{H}$ therefore necessarily differs from those in Ref.\cite{WISEMAN2000245}, in a way which we proceed to describe.

The characteristic polynomial of $LL^{\dagger}$ is given by $(\lambda^{2}-2\lambda +R^{2})^{2}$ (one can get this from the fact that $LL^{\dagger}-\lambda 1_{4}$ is a block diagonal matrix with invertible blocks), with roots $g^{2}(1\pm \vert T\vert)$ having geometric multiplicity 2. The $g^{2}(1+\vert T\vert)$ subspace ($g^{2}(1-\vert T\vert)$ subspace) is spanned by the first (last) two columns of
\begin{equation}
W=\begin{pmatrix}
{e^{i\theta_{T}}\over \sqrt{2}}&0&-{e^{i\theta_{T}}\over \sqrt{2}}&0\\
{1\over \sqrt{2}}&0&{1\over \sqrt{2}}&0\\
0&e^{i\theta_{T}}\sqrt{{1+\vert T\vert\over 2}}&0&-e^{i\theta_{T}}\sqrt{{1-\vert T\vert\over 2}} \\
0&\sqrt{{1-\vert T\vert\over 2}}&0&\sqrt{{1+\vert T\vert\over 2}}
\end{pmatrix}
\end{equation}
which is evidently unitary. From (\ref{eqn:phicov}) one obtains the $8\times 8$ covariance matrix $\Sigma_{\ket{\Phi_{L}}}=\Xi D\Xi^{\intercal}$ with $D={1\over 2}\left( \nu^{-}_{+}\mathbb{I}_{2} \oplus \nu^{-}_{-}\mathbb{I}_{2}\oplus\nu^{+}_{+}\mathbb{I}_{2}\oplus\nu^{+}_{-}\mathbb{I}_{2} \right)$, where $\nu^{\pm}_{+}:= e^{\pm 2g\sqrt{1+\vert T\vert}}$ and $\nu^{\pm}_{-}:= e^{\pm 2g\sqrt{1-\vert T\vert}}$ (see Appendix \ref{sec:nat}).

\begin{figure}[t]
    \includegraphics[scale=0.5]{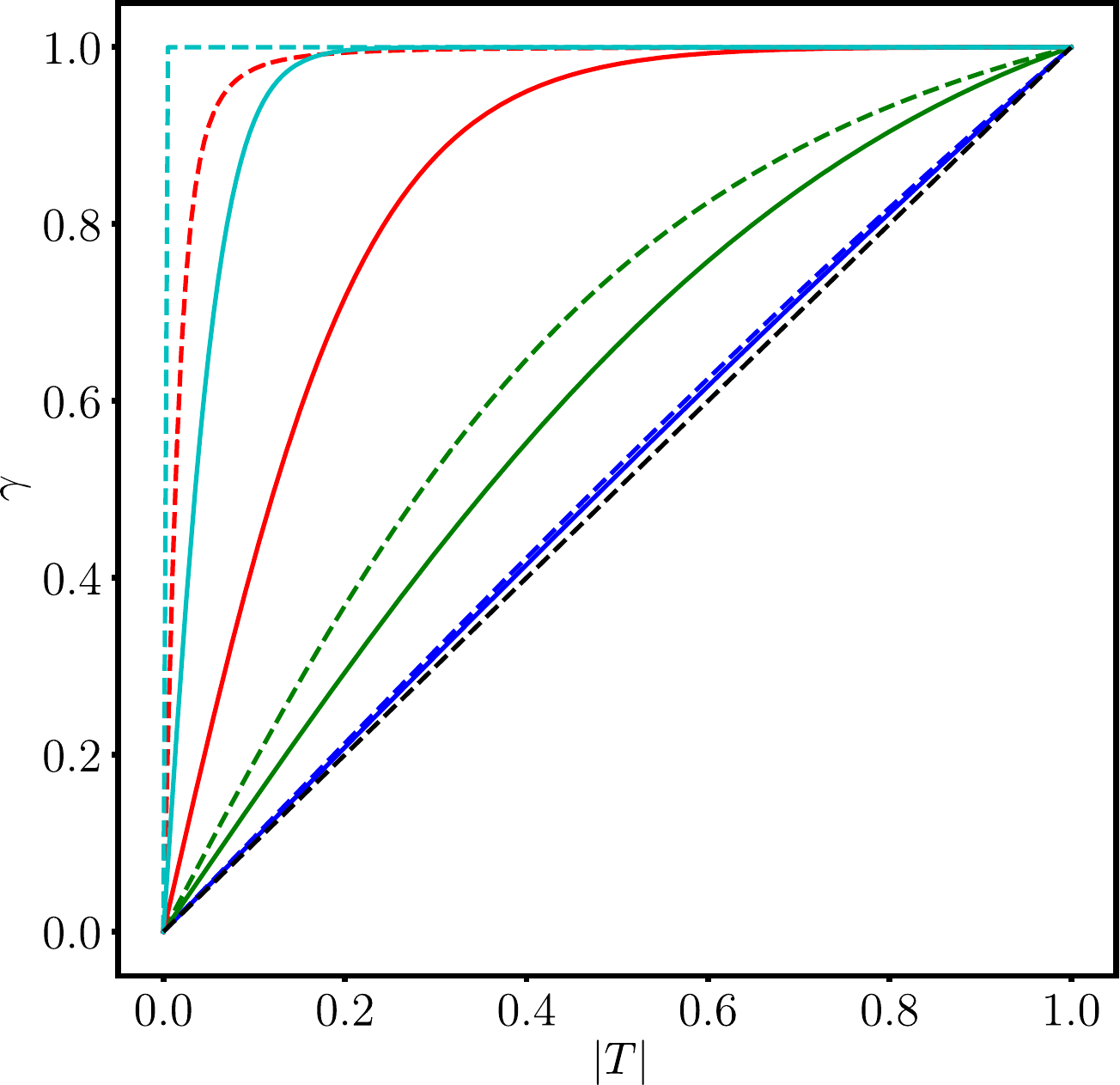}
    \caption{(Solid lines) Normalized first-order coherence for $e^{g} \in 0.37, 1.28, 4.48, 15.64$ for $\vert T\vert \in [0, 1]$
in the absence of phase shifts. Increasing concavity of curves indicates higher $g$. (Dashed lines)
Normalized first-order coherence from Ref.\cite{WISEMAN2000245} at the same values of $g$ and absence of phase shifts. Dashed black line is $\gamma = \vert T\vert$, lower bounding
all curves due to Cauchy-Schwarz inequality.}
    \label{fig:f1}
\end{figure}

Note that allowing for path identity has resulted in the  parameter $T$, which governs the idler indistinguishability, to appear in the symplectic eigenvalues of the covariance matrix. In other words, nonzero $T$ implies renormalization of the squeezing strengths in CV quadratures defined by $\Xi$, which occurs due to the $A_{S}$ and $B_{I}$ mode coupling in $\mathcal{H}$ and does not occur if the NL1 and NL2 downcoversions are assumed to occur sequentially.  The final step of the ZWM experiment implements a beamsplitter $U_{\text{BS}}$ on the signal modes $A_{S}$, $B_{S}$ that incorporates a phase shift $\phi_{S}$ due to possible differences in the signal paths. We model this beamsplitter by $U_{\text{BS}}a^{\dagger}U_{\text{BS}}^{\dagger}= a^{\dagger}Y$ where the unitary $Y$ is given by
\begin{align}
Y&:= {1\over \sqrt{2}}\begin{pmatrix}
1& ie^{i\phi_{S}} \\ i & e^{i\phi_{S}}
\end{pmatrix} \oplus 1_{2} .
\label{eqn:bs}
\end{align} 
The images of the modes $A_{S}$ and $B_{S}$ under the transformation $Y$ are the modes which are detected in Fig. \ref{fig:f1}. Consistent with the transformation of the creation operators in (\ref{eqn:bs}), which shows how the phases in the squeezed state network change under $\ket{\Phi_{L}}\mapsto U_{\text{BS}}\ket{\Phi_{L}}$, the transformation of the covariance matrix $\Sigma_{\ket{\Phi_{L}}}$ is given by 
\begin{align}
\Sigma_{\ket{\Phi_{L}}} &\mapsto \Sigma_{U_{\text{BS}}\ket{\Phi_{L}}} = O_{\text{BS}}^{T} \Sigma_{\ket{\Phi_{L}}} O_{\text{BS}} \nonumber \\
O_{\text{BS}} &:=  \begin{pmatrix}\text{Re}Y^{\intercal}&0&\text{Im}Y^{\intercal}&0\\
0&1_{2}&0&0 \\
-\text{Im}Y^{\intercal}&0&\text{Re}Y^{\intercal}&0\\
0&0&0&1_{2}\end{pmatrix}.
\label{eqn:jkjk}
\end{align}
Multiplication of the 8$\times$8 matrices in (\ref{eqn:jkjk}) and using the formula (\ref{eqn:photocurr}) for the expected photon number produces the following nonperturbative formulas for the photocurrents
\begin{align}
\langle a_{S_{1}}^{\dagger}a_{S_{1}} \rangle_{U_{\text{BS}}\ket{\Phi_{L}}}&= {\cosh(2g\sqrt{1+\vert T\vert})\over 4}\left(1-\sin(\phi_{S}+\theta_{T})\right) \nonumber \\
&{}+ {\cosh(2g\sqrt{1-\vert T\vert})\over 4}\left(1+\sin(\phi_{S}+\theta_{T})\right) -{1\over 2} \nonumber \\
&\overset{g\rightarrow 0}{=} g^{2}\left( 1-\vert T\vert \sin(\phi_{S}+\theta_{T}) \right) + o(g^{2}) \nonumber \\
\langle a_{S_{2}}^{\dagger}a_{S_{2}} \rangle_{U_{\text{BS}}\ket{\Phi_{L}}} &= {\cosh(2g\sqrt{1+\vert T\vert})\over 4}\left(1+\sin(\phi_{S}+\theta_{T})\right) \nonumber \\
&{} + {\cosh(2g\sqrt{1-\vert T\vert})\over 4}\left(1-\sin(\phi_{S}+\theta_{T})\right) -{1\over 2}\nonumber \\&\overset{g\rightarrow 0}{=} g^{2}\left(1+\vert T\vert \sin(\phi_{S}+\theta_{T}) \right) + o(g^{2}).
\label{eqn:uyuy}
\end{align} 
These formulas are the main result of the present work. They can be used to obtain all quantities of interest that are functions of the first moments of the photocurrent signals. As is clear from (\ref{eqn:uyuy}), when $g\rightarrow 0$ they reduce to the perturbative expressions given by ZWM \cite{RevModPhys.94.025007}, with the MacLaurin series starting at $O(g^{2})$. Note that for $T=0$, the effective SPDC elements are completely uncoupled and the sum of the energies of the signal modes is $2\sinh^{2}g$, exactly half of the total energy.

It is important to note that we do not consider the two downconversions as temporally separated dynamics, as was done in the nonperturbative analyses of Refs.\cite{WISEMAN2000245,Kolobov_2017,Miller2021versatilesuper}. The fact that: 1. the modes $A_{I}$ and $B_{I}$ are not perfectly distinguishable and, 2. the downcoversion occurs from coherently pumped emitters, imply that the parameter $T$ renormalizes the bare squeezing parameter $g$ in the resulting four-mode state. The maximum visibility of the photocurrent can be described by the modulus of the normalized first-order coherence function 
\begin{equation}
\gamma := {\vert\langle a_{S_{1}}^{\dagger}a_{S_{2}} \rangle_{U_{\text{BS}}\ket{\Phi_{L}}} \vert \over \sqrt{  \langle a_{S_{1}}^{\dagger}a_{S_{1}}\rangle_{U_{\text{BS}}\ket{\Phi_{L}}}\langle a_{S_{2}}^{\dagger}a_{S_{2}}\rangle_{U_{\text{BS}}\ket{\Phi_{L}}}}}
\label{eqn:gamm}
\end{equation} 
where the numerator can be obtained from the covariance matrix via
\begin{align}
4\vert\langle a_{S_{1}}^{\dagger}a_{S_{2}} \rangle_{U_{\text{BS}}\ket{\Phi_{L}}} \vert^{2} &=\left(  (\Sigma_{U_{\text{BS}}\ket{\Phi_{L}}})_{1,2}+(\Sigma_{U_{\text{BS}}\ket{\Phi_{L}}})_{5,6} \right)^{2}\nonumber \\
&{}+\left(  (\Sigma_{U_{\text{BS}}\ket{\Phi_{L}}})_{1,6}-(\Sigma_{U_{\text{BS}}\ket{\Phi_{L}}})_{2,5} \right)^{2}  .
\label{eqn:genvis}
\end{align}

Deviation of (\ref{eqn:gamm}) from linear dependence on $\vert T\vert$ is the hallmark of ZWM setup in high-gain regime \cite{WISEMAN2000245,Kolobov_2017,BELINSKY1992303}. Evaluating (\ref{eqn:gamm}) at $\theta_{T}=0$ and $\phi_{S}=0$ gives the result
\begin{equation}
\gamma \big\vert_{\theta_{T},\phi_{S}=0} = {\cosh(2g\sqrt{1+\vert T\vert})-\cosh(2g\sqrt{1-\vert T\vert}) \over \cosh(2g\sqrt{1+\vert T\vert}) + \cosh(2g\sqrt{1-\vert T\vert}) -2}
\label{eqn:analgamm}
\end{equation}
which is plotted in Fig.\ref{fig:f1}. The interpolation of the normalized first-order coherence function between linear
dependence on $\vert T\vert$ in the single-photon regime (i.e., the regime where low-order perturbation
theory in $g$ is valid) to independence from $T$ at unit value in the high-intensity regime is a
known phenomenon which is described by (\ref{eqn:analgamm}). However, note in Fig.\ref{fig:f1} that the result of Ref.\cite{WISEMAN2000245} given by
\begin{equation}
\gamma^{\text{(Ref.[12])}} \big\vert_{\theta_{T},\phi_{S}=0} ={\vert T\vert \cosh g\over \sqrt{1+\vert T\vert^{2}\sinh^{2}g}},
\label{eqn:ppp}
\end{equation}
arising from using factorized dynamics to describe the ZWM experiment, overestimates the normalized first-order coherence function. This is due to the fact that their analysis does not treat the
renormalization of squeezing that occurs due to the path indistinguishability, as is evident from the hyperbolic functions in (\ref{eqn:ppp}) which depend oly on the bare squeezing parameter $g$.

As quadrature squeezing levels increase toward 10 dB \cite{PhysRevLett.100.033602} for applications ranging from CV one-way quantum computing \cite{PhysRevA.79.062318} to gravitational wave detection \cite{PhysRevLett.126.041102}, it seems natural that the ZWM effect will be utilized at higher squeezing levels and downconversion rates, i.e., beyond the setting well-described by dynamics in the single-photon mode occupation subspace. Squeezed state networks $\ket{\Phi_{L}}$ are Glauber non-classical \cite{ZOU1991351}, which in turn implies improved performance for certain CV quantum sensing tasks \cite{PhysRevLett.122.040503}. Under a sequential model of the ZWM effect, it is already known that by placing an unknown phase shift $\theta$ on the $B_{I}$ mode between NL1 and NL2 in Fig.\ref{fig:fd}a to form a ZWM interferometer, the intensity difference of $A_{S}$ and $B_{S}$ provides a method of moments estimate of $\theta$ with error below the standard quantum limit \cite{Miller2021versatilesuper,PhysRevA.92.013832}. To determine the phase estimation precision achievable in the ZWM interferometer under the model of the present work, we calculate the quantum Fisher information (QFI) for a parametrized squeezed state network $\ket{\Phi_{L}(\theta)}$ seeded by a  coherent state $\ket{\beta}$ in the $A_{S}$ mode with intensity much greater than the energy due to quadrature noise, i.e., $\vert \beta\vert^{2}\gg e^{2g\sqrt{2}}$ (see Appendix \ref{sec:qfi} for full definition of probe state and QFI analysis). This distribution of energy is relevant to a realistic, high-precision phase estimation scenario. Using $F(\theta)=-2\partial_{\xi}^{2}\vert \langle \Phi_{L}(\theta)\vert\Phi_{L}(\xi)\rangle\vert^{2}\big\vert_{\xi=\theta}$ for the QFI, we obtain:

\begin{align}
F(\theta) &= {\beta^{2}\over 2}\left( \cosh g\sqrt{1+\vert T\vert} - \cosh g\sqrt{1-\vert T\vert} \right)^{2}\nonumber \\
&{} \left( \left( e^{-2g\sqrt{1-\vert T\vert}}+e^{-2g\sqrt{1+\vert T\vert}}\right)\cos^{2}\theta \right. \nonumber \\
&{}+ \left. \left( e^{2g\sqrt{1-\vert T\vert}}+e^{2g\sqrt{1+\vert T\vert}}\right)\sin^{2}\theta \right)\nonumber \\
&{} +o(\beta^{2}).
\label{eqn:bigfish}
\end{align}

Note that the QFI depends on $\theta$ because, unlike Mach-Zehnder or SU(1,1) interferometry, the parameter is not imprinted by $e^{-i\theta A}\ket{\psi}$ for unparametrized $\ket{\psi}$ and $A=A^{\dagger}$. Unlike previous analyses, one can now clearly see from (\ref{eqn:bigfish}) that the quality of the path identity operation, i.e., the closeness of $\vert T\vert$ to $1$, determines the sensitivity that is possible using undetected photons, which is consistent with physical demand that the sensitivity must vanish for $\vert T\vert=0$. For $\vert T\vert=1$, one sees that at the optimal angle $\theta=\pi/2$, the QFI scales as the product of the seed laser intensity $O(\beta^{2})$ with a factor scaling as the square of the downconverted photon intensity $O(e^{4\sqrt{2}g})$. This scaling shows that the phase sensitivity achievable in an SU(1,1) interferometer \cite{PhysRevA.33.4033} with coherent state seed port \cite{Plick_2010,PhysRevA.86.023844,Manceau_2017} or an SPDC/beamsplitter combined interferometer with coherent state seed port \cite{PhysRevA.87.023825,10.1063/5.0004873}
 is also achievable in a ZWM interferometer, where the photons being detected never propagated through the phase shift $\theta$. Formula (\ref{eqn:bigfish}) therefore indicates a tradeoff between the optimal sensitivity and the quality of path identity implementation.

\section{Discussion}
For future directions, we expect that by combining several pulsed SPDC and optical parametric oscillator sources with varying pulse delays, a fiber-based, time-multiplexed demonstration of the non-perturbative ZWM effect is possible, with potential applications including linear optical circuit characterization, quantum computation, and distributed quantum sensing. Finally, we note that states of the form (\ref{eqn:genstate}) also appear as effective descriptions as ground states of homogeneous systems of massive interacting bosons. Microscopically, a Bose-Einstein condensed zero-momentum mode serves as the analogue of the continuous-wave optical pump, and the two-body interaction implements the downconversion into opposite momentum modes. For example, the Bogoliubov ground state of the weakly-interacting Bose gas \cite{liebseiringer} and the Valatin-Butler wavefunction describing strongly-interacting boson systems \cite{Valatin1958} have this form. Number-conserving versions of these squeezed state networks have also been studied \cite{Leggett_2003,PhysRevA.67.033608,PhysRevA.93.033623}. Therefore, advances in methods for controlling the interaction strength could lead to a variety of ZWM effect analogues in the bosonic matter-wave setting.

\acknowledgements
The authors thank R. Newell, P. Milonni and R. Wenzel for relevant discussions and acknowledge the LDRD program at Los Alamos National Laboratory. Los Alamos National Laboratory is managed by Triad National Security, LLC, for the National Nuclear Security Administration of the U.S. Department of Energy under Contract No. 89233218CNA000001.
\onecolumngrid
\bibliography{np.bib}

\onecolumngrid
\appendix
\section{\label{sec:gs}Gaussian states}

We provide some background on the canonical formalism of CV Gaussian states which closely follows the notational conventions of Ref.\cite{holevo}, except we use ``$qp$-order'' of the row vector of canonical quadrature operators on $M$ CV modes which is also used in Refs.\cite{PhysRevA.37.3028,serafini}
\begin{equation}
\mathcal{R}=\begin{pmatrix}
q_{1} & \ldots & q_{M} & p_{1} & \ldots & p_{M}
\end{pmatrix}
\end{equation}
where for any column vectors $z,z'\in\mathbb{R}^{2M}$, $[\mathcal{R}z,\mathcal{R}z']=i\Delta(z,z')$ with $\Delta$  the symplectic form on $\mathbb{R}^{2M}$. We conflate the symbol $\Delta$ with its matrix, and use $\Delta =\begin{pmatrix}
0&1_{M}\\-1_{M}&0
\end{pmatrix}$ where $1_{M}$ is the $M\times M$ identity. We use $\dagger$ for the adjoint involution (Hermitian conjugate), $\intercal$ for matrix transpose, and a bar for complex conjugation.

The $M$-mode vacuum is $\ket{\text{VAC}}:=\ket{0}^{\otimes M}$. The row vector $a=\begin{pmatrix}
a_{1} & \ldots & a_{M} \end{pmatrix}$ of annihilation operators, along with its corresponding creation operators, are obtained from a linear transformation of the canonical quadrature operators
\begin{equation}
A:= (a,a^{\dagger}) = \mathcal{R}{1\over \sqrt{2}}\begin{pmatrix}
1_{M}&1_{M}\\i1_{M}& -i1_{M}
\end{pmatrix}.
\end{equation}
We will call a unitary $U$ generated by a (self-adjoint) homogeneous polynomial of order 2 in the creation and annihilation operators a Gaussian unitary. A Gaussian unitary acts on the canonical variables as
\begin{equation}
U^{\dagger}\mathcal{R}U=\mathcal{R}T_{U}
\label{eqn:gu}
\end{equation}
where $T_{U}$ is a $2M\times 2M$ symplectic matrix with respect to $\Delta$. Any pure, zero-mean Gaussian state has the form $U\ket{\text{VAC}}$ for Gaussian unitary $U$, and is completely specified by its covariance matrix
\begin{equation}
\Sigma_{U\ket{\text{VAC}}}:= \langle \mathcal{R}^{\intercal}\circ \mathcal{R}\rangle_{U\ket{\text{VAC}}} = {1\over 2}T_{U}^{\intercal}T_{U} 
\end{equation}
where for operators $A,B$ the Jordan product is $A\circ B:= {1\over 2}AB+{1\over 2}BA$. A general pure Gaussian state $\ket{\psi}$ is specified by its mean vector $m_{\ket{\psi}}:=\langle \mathcal{R}\rangle_{\ket{\psi}}$ and its covariance matrix, the definition of the latter now being $\langle (\mathcal{R}-m_{\ket{\psi}})^{\intercal}\circ (\mathcal{R}-m_{\ket{\psi}})\rangle_{\ket{\psi}}$. A general pure Gaussian state can be obtained from vacuum by applying a Gaussian unitary $U$ acting on $\mathcal{R}$ as in (\ref{eqn:gu}) to the Fock vacuum, followed by a displacement unitary $D(z):= e^{i\mathcal{R}z}$ for $z\in \mathbb{R}^{2M}$.

\section{\label{sec:nat}Necessity of Autonne-Takagi diagonalization; explicit covariance matrix $\Sigma_{\ket{\Phi_{L}}}$}

It would be desirable to obtain the covariance matrix $\Sigma_{\ket{\Phi_{L}}}$ by straightforwardly computing the adjoint action of the squeezing operator on the creation and annihilation operators
\begin{equation}
A':= e^{-{1\over 2}a^{\dagger}La^{\dagger \intercal} + h.c.}Ae^{{1\over 2}a^{\dagger}La^{\dagger \intercal} - h.c.}
\end{equation}
using the Baker-Campbell-Hausdorff formula. Although a closed form is readily obtained in special cases in which $L$ is periodic with small period, in general one must be satisfied with the series
\begin{equation}
A' =  \begin{pmatrix}\cosh(L)&- \sinh(L)\\- \sinh(L)&\cosh(L)\end{pmatrix}A.\end{equation}  The most direct route to $\Sigma_{\ket{\Phi_{L}}}$, then, is the Autonne-Takagi diagonalization  of the squeezing matrix $L$ \cite{horn}.

From (3) and (5) of the main text, one can write explicitly the $8\times 8$ covariance matrix $\Sigma_{\ket{\Phi_{L}}}=\Xi D\Xi^{\intercal}$ with $D={1\over 2}\left( \nu^{-}_{+}\mathbb{I}_{2} \oplus \nu^{-}_{-}\mathbb{I}_{2}\oplus\nu^{+}_{+}\mathbb{I}_{2}\oplus\nu^{+}_{-}\mathbb{I}_{2} \right)$, where $\nu^{\pm}_{+}:= e^{\pm 2g\sqrt{1+\vert T\vert}}$ and $\nu^{\pm}_{-}:= e^{\pm 2g\sqrt{1-\vert T\vert}}$. Calling $C_{j}$ the $j$-th column of $\Sigma_{\ket{\Phi_{L}}}$, one finds that
\begin{align}
&{} C_{1}={1\over 4}\begin{pmatrix}
\left(\nu^{+}_{-}+\nu^{+}_{+}\right)\sin^{2}\theta_{T}+\left(\nu^{-}_{+}+\nu^{-}_{-}\right)\cos^{2}\theta_{T}\\
\left( \nu^{-}_{+}-\nu^{-}_{-}\right)\cos \theta_{T} \\
0\\
0\\
{1\over 2}\left(  \nu^{+}_{-}+\nu^{+}_{+}-\nu^{-}_{+}-\nu^{-}_{-}\right)\sin 2\theta_{T} \\
\left( \nu^{+}_{+}-\nu^{+}_{-}\right)\sin \theta_{T} \\
0\\
0
\end{pmatrix} \; , \; 
C_{2}={1\over 4}\begin{pmatrix}
 \lbrace{ \rbrace}\\
 \nu^{-}_{+}+\nu^{-}_{-} \\
 0\\
 0\\
 \left( \nu^{-}_{-}-\nu^{-}_{+} \right)\sin\theta_{T}\\
 0\\
 0\\
 0\\
\end{pmatrix} \nonumber \\
&{} C_{3} = \begin{pmatrix}
\lbrace{ \rbrace}\\
\lbrace{ \rbrace}\\
\left(1-\vert T\vert \right)\left( \nu^{+}_{-}\sin^{2}\theta_{T} + \nu^{-}_{-}\cos^{2}\theta_{T}\right) + \left(1+\vert T\vert \right)\left( \nu^{+}_{+}\sin^{2}\theta_{T} + \nu^{-}_{+}\cos^{2}\theta_{T}\right) \\
\sqrt{1-\vert T\vert^{2}}\left( \nu^{-}_{+} -\nu^{-}_{-}\right)\cos\theta_{T} \\
0\\
0\\
{1-\vert T\vert\over 2}\left( \nu^{+}_{-} -\nu^{-}_{-} \right) \sin 2\theta_{T} + {1+\vert T\vert\over 2}\left( \nu^{+}_{+} -\nu^{-}_{+} \right) \sin 2\theta_{T} \\
\sqrt{1-\vert T\vert^{2}}\left( \nu^{+}_{+} -\nu^{+}_{-} \right) \sin\theta_{T}
\end{pmatrix}\nonumber \\
&{} C_{4} = \begin{pmatrix}
\lbrace{ \rbrace}\\
\lbrace{ \rbrace}\\
\lbrace{ \rbrace}\\
\left(1-\vert T\vert\right) \nu^{-}_{+} + \left(1+\vert T\vert\right) \nu^{-}_{-} \\
0\\
0\\
\sqrt{1-\vert T\vert^{2}}\left( \nu^{-}_{-} -\nu^{-}_{+} \right) \sin\theta_{T}\\
0
\end{pmatrix}\; , \; C_{5} = \begin{pmatrix}
\lbrace{ \rbrace}\\
\lbrace{ \rbrace}\\
\lbrace{ \rbrace}\\
\lbrace{ \rbrace}\\
\left( \nu^{+}_{+} +\nu^{+}_{-} \right)\cos^{2}\theta_{T}  + \left( \nu^{-}_{+} +\nu^{-}_{-} \right)\sin^{2}\theta_{T}\nonumber \\
\left( \nu^{+}_{+} -\nu^{+}_{-} \right)\cos\theta_{T}
\end{pmatrix}\nonumber \\
&{} C_{6} = \begin{pmatrix}
\lbrace{ \rbrace}\\
\lbrace{ \rbrace}\\
\lbrace{ \rbrace}\\
\lbrace{ \rbrace}\\
\lbrace{ \rbrace}\\
\nu^{+}_{-} +\nu^{+}_{+} \\
0\\
0
\end{pmatrix}\; , \; 
 C_{7} = \begin{pmatrix}
\lbrace{ \rbrace}\\
\lbrace{ \rbrace}\\
\lbrace{ \rbrace}\\
\lbrace{ \rbrace}\\
\lbrace{ \rbrace}\\
\lbrace{ \rbrace}\\
\left(1-\vert T\vert \right)\left( \nu^{+}_{-}\cos^{2}\theta_{T} + \nu^{-}_{-}\sin^{2}\theta_{T}\right) + \left(1+\vert T\vert \right)\left( \nu^{+}_{+}\cos^{2}\theta_{T} + \nu^{-}_{+}\sin^{2}\theta_{T}\right) \\
\sqrt{1-\vert T\vert^{2}}\left( \nu^{+}_{+} -\nu^{+}_{-} \right)\cos\theta_{T}
\end{pmatrix}\nonumber \\
&{} C_{8}= \begin{pmatrix}
\lbrace{ \rbrace}\\
\lbrace{ \rbrace}\\
\lbrace{ \rbrace}\\
\lbrace{ \rbrace}\\
\lbrace{ \rbrace}\\
\lbrace{ \rbrace}\\
\lbrace{ \rbrace}\\
\left(1-\vert T\vert\right)\nu^{+}_{+} + \left( 1+\vert T\vert \right)\nu^{+}_{-}
\end{pmatrix}
\end{align}
where blank entries are specified by previous columns because $\Sigma_{\ket{\Phi_{L}}}^{\intercal}=\Sigma_{\ket{\Phi_{L}}}$.

\section{\label{sec:qfi}Quantum Fisher information for $\ket{\Phi_{L}(\theta)}$ probe}

Due to the fact that a phase shift $\theta$ on the $B_{I}$ mode is equivalent to taking a complex transmissivity parameter $T$, the probe state is
\begin{equation}
\ket{\Phi_{L}(\theta)} := e^{\mathcal{H}}\ket{\beta}_{A_{S}}\ket{0}_{B_{S}}\ket{0}_{B_{I}}\ket{0}_{X}
\end{equation}
where we take $\beta\in\mathbb{R}$ and $\mathcal{H} := {1\over 2}a^{\dagger}La^{\dagger \intercal} - h.c.$ where $L$ is the same as in the main text except with $\theta_{T}\mapsto \theta$. We define the characteristic function of an $M$-mode CV quantum state $\rho$ by
\begin{equation}
\chi_{\rho}(z):=\text{tr}\rho e^{i\mathcal{R}z}
\end{equation}
which fully specifies the state $\rho$ via the inverse relation
\begin{equation}
\rho={1\over (2\pi)^{M}}\int d^{2M}z \, \chi_{\rho}(z)e^{-i\mathcal{R}z}.
\end{equation}
Using this relation and the quantum Fisher information on the pure state manifold $\lbrace \ket{\Phi_{L}(\theta)} : \theta \in [0,2\pi)\rbrace$ given by $F(\theta)=-2\partial_{\xi}^{2}\vert \langle \Phi_{L}(\theta)\vert\Phi_{L}(\xi)\rangle\vert^{2}\big\vert_{\xi=\theta}$ gives the explicit formula for $F(\theta)$ in terms of mean vector and covariance \cite{PhysRevA.85.010101,saf}
\begin{equation}
F(\theta)={dm_{\ket{\Phi_{L}(\theta)}}\over d\theta}\Sigma_{\ket{\Phi_{L}(\theta)}}^{-1}\left(  {dm_{\ket{\Phi_{L}(\theta)}}\over d\theta}\right)^{\intercal} +{1\over 4}\text{tr} \left[ \left( \Sigma_{\ket{\Phi_{L}(\theta)}}^{-1}{d\Sigma_{\ket{\Phi_{L}(\theta)}}\over d\theta} \right)^{2}\right].
\label{eqn:fishfish}
\end{equation}
When the $A_{S}$ mode is seeded by a coherent state $\ket{\beta}$ with energy $\vert \beta\vert^{2}$ much larger than the energy due to quadrature noise from SPDC, it is only necessary to keep track of the first term in (\ref{eqn:fishfish}) because it scales as $O(\beta^{2})$ whereas the second term does not depend on $\beta$. From the diagonalization $\Sigma_{\ket{\Phi_{L}(\theta)}}=\Xi D\Xi^{\intercal}$, it is straightforward to compute $\Sigma_{\ket{\Phi_{L}(\theta)}}^{-1}$. The mean vector $m_{\ket{\Phi_{L}(\theta)}}$ can be obtained by again appealing to the Autonne-Takagi diagonalization. Note that $\ket{\beta}_{A_{S}}=e^{-{\beta^{2}\over 2}}e^{\beta a_{A_{S}}^{\dagger}}\ket{0}_{A_{S}}$, so to determine the equivalent displacement $D(z)$ satisfying $D(z)e^{\mathcal{H}}\ket{0}_{A_{S}}\ket{0}_{B_{S}}\ket{0}_{B_{I}}\ket{0}_{X} = e^{\mathcal{H}}\ket{\beta}_{A_{S}}\ket{0}_{B_{S}}\ket{0}_{B_{I}}\ket{0}_{X}$, one computes (with our convention $(a_{1},a_{2},a_{3},a_{4}):= (a_{A_{S}},a_{B_{S}},a_{B_{I}},a_{X})$)
\begin{align}
e^{\mathcal{H}}a_{1}^{\dagger}e^{-\mathcal{H}} &= \mathcal{W}^{\dagger}e^{{1\over 2}\sum_{j=1}^{4}\left[\lambda_{j}a_{j}^{\dagger 2} - h.c.\right]}\mathcal{W}a^{\dagger}_{1}\mathcal{W}^{\dagger}e^{-{1\over 2}\sum_{j=1}^{4}\left[\lambda_{j}a_{j}^{\dagger 2} - h.c.\right]}\mathcal{W}\nonumber \\
&= \mathcal{W}^{\dagger}e^{{1\over 2}\sum_{j=1}^{4}\left[\lambda_{j}a_{j}^{\dagger 2} - h.c.\right]}\left(a^{\dagger}W^{\dagger}\right)_{1}e^{-{1\over 2}\sum_{j=1}^{4}\left[\lambda_{j}a_{j}^{\dagger 2} - h.c.\right]}\mathcal{W}\nonumber \\
&= \mathcal{W}^{\dagger}e^{{1\over 2}\sum_{j=1}^{4}\left[\lambda_{j}a_{j}^{\dagger 2} - h.c.\right]}\left({a_{A_{S}}^{\dagger}e^{-i\theta}-a_{B_{I}}^{\dagger}e^{-i\theta}\over \sqrt{2}}\right)e^{-{1\over 2}\sum_{j=1}^{4}\left[\lambda_{j}a_{j}^{\dagger 2} - h.c.\right]}\mathcal{W}\nonumber \\
&= {e^{-i\theta}\over\sqrt{2}}\mathcal{W}^{\dagger} \left( \cosh g\sqrt{1+\vert T\vert}a_{A_{S}}^{\dagger}-\sinh g\sqrt{1+\vert T\vert}a_{A_{S}} - \cosh g\sqrt{1-\vert T\vert} a_{B_{I}}^{\dagger} + \sinh g\sqrt{1-\vert T\vert}a_{B_{I}}\right) \mathcal{W}\nonumber \\
&= {1\over 2}\left( \cosh g\sqrt{1+\vert T\vert}(a_{A_{S}}^{\dagger}+e^{-i\theta}a_{B_{S}}^{\dagger})-\sinh g\sqrt{1+\vert T\vert}(a_{A_{S}}-a_{B_{I}})\right. \nonumber \\
&{} \left. - \cosh g\sqrt{1-\vert T\vert}(e^{-i\theta}a_{B_{S}}^{\dagger}-a_{A_{S}}^{\dagger}) + \sinh g\sqrt{1-\vert T\vert}(\sqrt{1+\vert T\vert}a_{B_{S}}-\sqrt{1-\vert T\vert}a_{X}) \right)
\end{align}
where the fourth line uses the adjoint action of the unitary squeezing operator on the creation operator, which follows from $e^{{1\over 2}(ra^{\dagger 2}-ra^{2})}qe^{-{1\over 2}(ra^{\dagger 2}-ra^{2})} = e^{-r}q$. The fifth line uses $\mathcal{W}^{\dagger}a^{\dagger}\mathcal{W}=a^{\dagger}W$. By calculating also $e^{\mathcal{H}}a_{1}e^{-\mathcal{H}}$, one can determine the displacement $D(z)$ such that the probe state can be written $D(z)e^{\mathcal{H}}\ket{\text{VAC}}_{A_{S}B_{S}B_{I}X}$ and, thereby, the mean vector $m_{\ket{\Phi_{L}(\theta)}}$. One finds that
\begin{equation}
{dm_{\ket{\Phi_{L}(\theta)}}\over d\theta}=-{\beta \over \sqrt{2}}\left( \cosh g\sqrt{1+\vert T\vert} -\cosh g\sqrt{1-\vert T\vert} \right)\begin{pmatrix}
0\\
\sin\theta\\
0 \\
0\\
0\\
\cos \theta\\
0\\
0\\
\end{pmatrix}.
\end{equation}
The first term of (\ref{eqn:fishfish}) is
\begin{align}
{dm_{\ket{\Phi_{L}(\theta)}}\over d\theta} \Xi D^{-1}\Xi^{\intercal}{dm_{\ket{\Phi_{L}(\theta)}}\over d\theta} &= {\beta^{2}\over 2}\left( \cosh g\sqrt{1+\vert T\vert} - \cosh g\sqrt{1-\vert T\vert} \right)^{2}\nonumber \\
&{} \left( \left( e^{-2g\sqrt{1-\vert T\vert}}+e^{-2g\sqrt{1+\vert T\vert}}\right)\cos^{2}\theta + \left( e^{2g\sqrt{1-\vert T\vert}}+e^{2g\sqrt{1+\vert T\vert}}\right)\sin^{2}\theta \right)
\end{align}
and the second term of (\ref{eqn:fishfish}) is $o(\beta^{2})$.

\end{document}